\documentclass[titlepage,12pt]{article}
\usepackage{hyperref}
\usepackage{cite}
\usepackage{amssymb,amsmath,color,graphics,amscd,epsf,indentfirst,amsfonts}
\usepackage{epsfig}
\usepackage{nicefrac}
\usepackage{scalefnt}
\usepackage[titles]{tocloft}
\usepackage{sectsty}
\allsectionsfont{\bf \scalefont{.7} \selectfont}
\subsectionfont{\bf \scalefont{.85} \it \selectfont}
\subsubsectionfont{\bf \scalefont{1} \it \selectfont}
\usepackage[T1]{fontenc}
\usepackage{lmodern}
\usepackage{bibspacing}

\def\blfootnote{\xdef\@thefnmark{}\@footnotetext}

\long\def\symbolfootnote[#1]#2{\begingroup%
\def\thefootnote{\fnsymbol{footnote}}\footnote[#1]{#2}\endgroup}

\newcommand{\oao}[2]{{#1\atopwithdelims[]#2}}

\makeatletter
\renewcommand{\@dotsep}{4.5}
\makeatother

\def\be{\begin{equation}}
\def\ee{\end{equation}}

\makeatletter
\def\@seccntformat#1{\csname the#1\endcsname.\quad}
\makeatother

\def\clock{{\count0=\time
           \divide\count0 60
           \ifnum\count0<10 0\fi\the\count0
           \multiply\count0 -60 \advance\count0 \time
           :\ifnum\count0<10 0\fi \the\count0
         }}
\newcommand{\timestamp}{{\small\vbox{\hbox{\tt\jobname.tex}
\hbox{\the\day/\the\month/\the\year, \clock}}}}

\def\NN{{\cal N}}
\def\OO{{\cal O}}

\def\SS{{\cal S}}

\newcommand{\beq}{\begin{equation}}
\newcommand{\eeq}{\end{equation}}
\newcommand{\ba}{\begin{array}}
\newcommand{\ea}{\end{array}}
\newcommand{\bea}{\begin{eqnarray}}
\newcommand{\eea}{\end{eqnarray}}

\newcommand{\Z}{\mathbb{Z}}

\newcommand{\R}{\mathbb{R}}

\newcommand{\tr}{\mathop{{\rm Tr}}}

\numberwithin{equation}{section}

\begin{document}

\begin{titlepage}
\begin{flushright}
CPHT-RR011.0209\\
\vskip -1cm
\end{flushright}
\vskip 1.3cm
\begin{center}
\font\titlerm=cmr10 scaled\magstep4
    \font\titlei=cmmi10 scaled\magstep4
    \font\titleis=cmmi7 scaled\magstep4
    \centerline{\titlerm
    Brane Dynamics and 3D Seiberg Duality}
      \vspace{0.4cm}
    \centerline{\titlerm
    on the Domain Walls of 4D \LARGE{$\NN=1$} SYM}
\vskip 1cm
{Adi Armoni$^\dagger$, Amit Giveon$^\ast$\symbolfootnote[4]{Permanent address:
Racah Institute of Physics, The Hebrew University, Jerusalem 91904, Israel},
Dan Isra\"el$^\ddagger$ and Vasilis Niarchos$^\natural$}\\
\vskip 0.5cm
\medskip
{\it $^{\dagger}$Department of Physics, Swansea University \\
Singleton Park, Swansea, SA2 8PP, UK}\\ {\tt a.armoni@swansea.ac.uk} \\
\smallskip
{\it $^{\ast}$EFI and Department of Physics, University of Chicago\\
5640 S.\ Ellis Av., Chicago, IL 60637, USA}\\
{\tt giveon@phys.huji.ac.il}\\
\smallskip
{\it $^{\ddagger}$GRECO, Institut d'Astrophysique de Paris,
98bis Bd Arago, 75014 Paris, France}\\
{\it Unit\'e mixte de Recherche 7095, CNRS -- Universit\'e Pierre et Marie Curie}\\
{\tt israel@iap.fr}\\
\smallskip
{\it $^\natural$Centre de Physique Th\'eorique, \'Ecole Polytechnique,
91128 Palaiseau, France}\\
{\it Unit\'e mixte de Recherche 7644, CNRS}\\
{\tt niarchos@cpht.polytechnique.fr}
\end{center}
\vskip .3cm
\centerline{\bf Abstract}

\baselineskip 15pt
%

\vskip .5cm \noindent
We study a three-dimensional $U(k)$ Yang-Mills Chern-Simons theory with 
adjoint matter preserving two supersymmetries. According to Acharya and Vafa, 
this theory describes the low-energy worldvolume dynamics of BPS domain walls 
in four-dimensional $\NN=1$ SYM theory. We demonstrate how to obtain the same 
theory in a brane configuration of type IIB string theory that contains threebranes and 
fivebranes. A combination of string and field theory techniques allows us to re-formulate 
some of the well-known properties of $\NN=1$ SYM domain walls in a geometric language 
and to postulate a Seiberg-like duality for the Acharya-Vafa theory. In the process, 
we obtain new information about the dynamics of branes in setups that preserve two 
supersymmetries. Using similar methods we also study other $\NN=1$ CS theories 
with extra matter in the adjoint and fundamental representations of the gauge group.

\vfill
\end{titlepage}\vfill\eject


{\it {\large In honor of Mikhail Shifman, on the occasion of his 60$^{\rm th}$ birthday.}}

\vspace{0.5cm}

\setcounter{equation}{0}
\pagestyle{empty}
\small
\vspace*{-0.7cm}
\normalsize
\pagestyle{plain}
\setcounter{page}{1}

\section{Introduction}
\label{sec:intro}

Brane configurations in string theory~\cite{Hanany:1996ie} have proven to be a
useful tool in studying the strong coupling regime of gauge theories with various
amounts of supersymmetry in diverse dimensions; see ref.~\cite{Giveon:1998sr} for
a review. In this paper we study a certain class of three-dimensional Yang--Mills Chern--Simons
(YM-CS) theories with two supersymmetries. Other YM-CS theories with larger amounts of
supersymmetry attracted recently a lot of attention due to their relation with the worldvolume
theory of M-theory membranes and the AdS$_4$/CFT$_3$ correspondence~\cite{Aharony:
2008ug}. An interesting by-product of this discussion was the realization
that some of these theories admit a Seiberg-type duality~\cite{Giveon:2008zn,Niarchos:2008jb,
Niarchos:2009aa,Amariti:2009rb}.

The theory that we mainly study here has a $U(k)$ gauge group and one $\NN=1$ 
adjoint scalar multiplet.  It can be described as an $\NN=2$ super-Yang--Mills (SYM) 
theory with an $\NN=1$ Chern--Simons interaction that reduces the amount of 
supersymmetry by half. The action of the theory is 
\beq
\label{avac}
\SS=\SS_{\NN=2 ~\textsc{SYM}}+\SS_{\NN=1 ~\textsc{CS}}
~,
\eeq
with
\begin{subequations}
\label{avafull}
\begin{align}
\label{avad}
\SS_{\NN=2~\textsc{SYM}}&=\frac{1}{4g^2_\textsc{ym}} \int \text{d}^3x ~
\tr \left( (D \phi)^2-F^2+i \bar \chi \displaystyle{\not}D \chi
+i \bar \psi \displaystyle{\not}D \psi+2i \bar \chi[\phi,\psi]
\right)\\
\label{avae}
\SS_{\NN=1~ \textsc{CS}}&=\frac{N}{4\pi} \int
 \tr \left( A\wedge \text{d}A+\tfrac{2}{3} A\wedge A\wedge A \right)
-\frac{N}{4\pi} \int \text{d}^3 x~\bar \chi \chi~.
\end{align}
\end{subequations}
The gauge field $A$ and the Majorana fermion $\chi$ form an $\NN=1$
vector multiplet ($F$ denotes the field strength of $A$). The real scalar field 
$\phi$ and the Majorana fermion $\psi$ form an $\NN=1$ scalar multiplet.

One motivation for studying this theory originates from the following observation.
In $\NN=1$ SYM theory in four dimensions, the $U(1)_R$ symmetry is broken
to ${\Z}_{2N}$ by the chiral anomaly. The ${\Z}_{2N}$ symmetry is further broken
spontaneously down to ${\Z}_2$ by gaugino condensation. As a result, there are 
BPS domain walls which interpolate between the various vacua of the theory
\cite{Dvali:1996xe}. Witten proposed \cite{Witten:1997ep} that the domain walls 
of $\NN=1$ SYM behave like D-branes for the QCD-string, as the QCD-string can 
end on them. In that case, one expects that there is a gauge theory living on the 
SYM walls. Later, it was argued by Acharya and Vafa (AV)~\cite{Acharya:2001dz}, 
using string theory, that the theory that lives on $k$ coincident domain walls of 
$\NN=1$ SYM is the three-dimensional $U(k)$ YM-CS gauge theory that appears
in eqs.~\eqref{avac}-\eqref{avafull}. Ref.~\cite{Acharya:2001dz} obtained this 
result from a large $N$ transition in a type IIA setup that will be reviewed in 
subsection~\ref{Tduality}.

\begin{figure}[t!]
\centerline{\includegraphics[width=6cm]{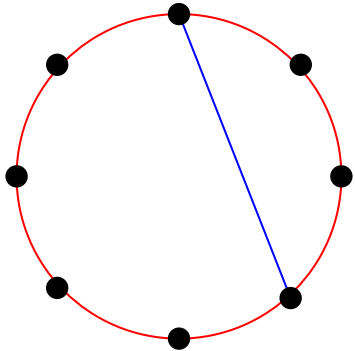}}\bf
\caption{\it The vacua and $k$-walls of $\NN=1$ SYM. The $k$-wall and the
$(N-k)$-anti-wall interpolate between the same vacua. The present example is
$N=8$: the depicted $3$-wall is equivalent to a $5$-anti-wall.}
\label{fig:vacua}
\end{figure}

The appearance of the AV theory in this context raises the following question
\cite{Armoni:2005sp}: since the `clockwise' interpolation between $k$ vacua
and the `anti-clockwise' interpolation between $N-k$ vacua are the same in 
$\NN=1$ SYM (see fig.~\ref{fig:vacua}), is it sensible to conclude that the $U(k)$ 
level $N$ AV theory is equivalent to the $U(N-k)$ level $N$ theory? Furthermore, 
what is the nature of this duality in three-dimensional field theory terms?

In this work, we will answer this question by proposing that the $U(k)$ and the
$U(N-k)$ theories form a pair of Seiberg dual theories. The two theories in this
pair flow to the same infrared (IR) theory and hence naturally describe the same 
$k$-wall (or $(N-k)$-anti-wall) bound state. At low energies, below the energy scale 
set by the gauge boson mass (coming from the CS interaction), the standard Yang--Mills 
kinetic terms can be dropped and both theories become $\NN=1$ CS theories
coupled to an $\NN=1$ scalar multiplet. We will argue for a duality that relates 
this pair of $\NN=1$ Chern--Simons-matter (CSM) theories. In the deep IR,
where both theories become topological, the equivalence reduces to a 
well-known level-rank duality between bosonic Chern--Simons theories.

Additional arguments for Seiberg duality in three dimensions will be provided 
by using a brane configuration in type IIB string theory, which is an $\NN=1$ deformation 
of the setup that was used in~\cite{Giveon:2008zn} to argue for Seiberg duality 
in $\NN=2$ CSM theories. The brane configuration consists of $k$ coincident 
D3-branes suspended between an NS5-brane and a $(1,N)$ fivebrane bound 
state (see sec.\ \ref{sec:HW} for a detailed description). After T-duality this setup
bears many similarities with the large-$N$ dual string theory background of 
\cite{Acharya:2001dz} but is not identical to it.

The study of this brane configuration will be doubly beneficial. On the one hand,
it provides an intuitive geometric reformulation of non-perturbative gauge
theory dynamics in three dimensions. On the other hand, we can use the 
field theory picture to learn more about brane dynamics in a setup that preserves only
two supersymmetries. We will see, in particular, how the $s$-rule of brane
dynamics and the brane creation (Hanany--Witten) effect work together
with a perturbatively generated potential for a pseudo-modulus that binds
suspended D3-branes into a bound state.

The main results of this paper are as follows. In section~\ref{sec:HWSYM}
we study the AV theory using a combination of string and field theory
techniques and argue that it admits a Seiberg-like duality for $k \leqslant N$. When
$k>N$, the $s$-rule dictates that supersymmetry is spontaneously broken
in agreement with field theory expectations from the $\NN=1$ SYM theory
in four dimensions. We evaluate the degeneracy $I_{k,N}$ of a $k$-wall by
lifting the brane configuration to M-theory. The result is
\beq
I_{k,N} = {N! \over  k! (N-k)!} ~,
\label{indexaa}
\eeq
in agreement with the field theory calculation of ref.~\cite{Acharya:2001dz}.
{From} the brane point of view, a particularly interesting part of the story is
how D3-branes (corresponding to the domain walls of $\NN=1$ SYM)
attract each other to form bound states with a tension given by the formula~\eqref{tensionaa}.

Two generalizations of the AV theory are discussed using brane techniques
in section~\ref{sec:general}. The first generalization considers the addition
of fundamental matter to the field content of the AV theory, and the second
the addition of extra matter in the adjoint in the presence of a tree-level
superpotential. We conclude in section~\ref{sec:conclusions} with a brief
summary of our results and a list of interesting questions and open problems.

\section{$\NN=1$ Chern--Simons-Matter Theories from Branes}
\label{sec:HW}

Supersymmetric gauge theories in diverse dimensions arise naturally,
as low-energy effective descriptions, in configurations of D-branes and
NS5-branes in type II string theory (see~\cite{Giveon:1998sr} for a review). 
In this section we revisit a configuration that realizes Chern--Simons-Matter 
theories with $\NN=1$ supersymmetry. For special values of the parameters 
that characterize the configuration we recover the AV CS theory which 
describes the low-energy dynamics on the domain walls of $\NN=1$ SYM 
theory in four dimensions.

\subsection{The brane setup of interest}

\begin{figure}[t!]
\centering
\includegraphics[height=5cm]{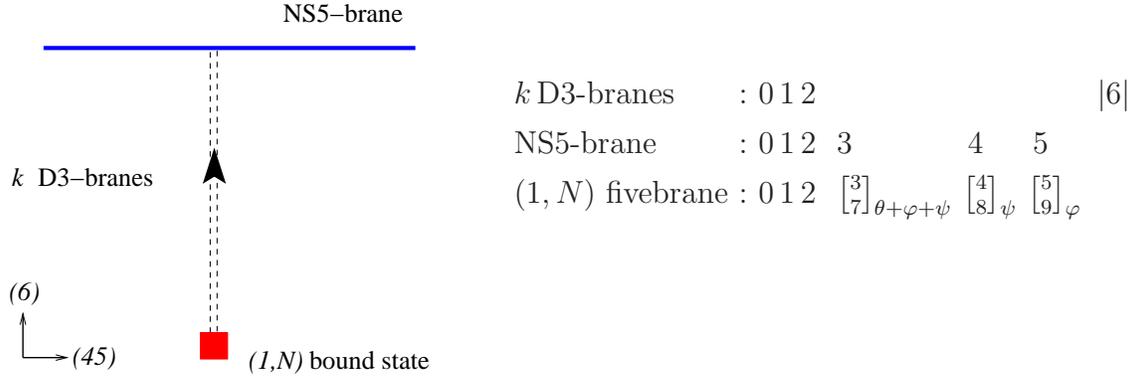}
\raisebox{3cm}{$\qquad  \begin{array}{lclllll}
k\, \text{D3-branes} & : &0\, 1\, 2\, &&&&\, |6|\\
\text{NS5-brane} & : & 0\, 1\, 2& \, 3&\, 4&\, 5&  \\
(1,N)\ \text{fivebrane} &  : & 0\, 1\, 2 &\, \oao{3}{7}_{\theta+\varphi+\psi} &\oao{4}{8}_\psi& \oao{5}{9}_{\varphi}
\end{array}$}\bf
\caption{\it  The brane configuration of interest. For generic angles $\psi,\varphi$
only two supercharges are conserved. The low-energy dynamics is described by
a three-dimensional $\mathcal{N}=1$ CSM theory.}
\label{sec2:orientN1}
\end{figure}

Consider a configuration of branes consisting of $k$ D3-branes, one NS5-brane
and a bound state of one NS5-brane and $N$ D5-branes, $i.e.$ a $(1,N)$
fivebrane, oriented as depicted in fig.~\ref{sec2:orientN1}.
The D3-branes are suspended between the fivebranes and have a finite
extent $L$ along the $x^6$ direction, a feature captured by the notation $|6|$.
The orientation of the $(1,N)$ bound state along the $(37)$, $(48)$ and $(59)$
planes is given by the angles $\theta$, $\psi$ and $\varphi$.\footnote{$\oao{i}{j}_\vartheta$
denotes that the brane is oriented along the $(ij)$ plane at an angle $\vartheta$
with respect to the axis $i$.}
The configuration preserves at least two supersymmetries for generic angles $\varphi$, $\psi$
provided $\theta$ obeys the following relation~\cite{Aharony:1997ju,Kitao:1998mf}:
\begin{equation}
\tan \theta = g_s N
~,
\label{sec2:thetavalue}
\end{equation}
where $g_s$ is the string coupling. 

The low-energy effective theory that describes the dynamics of this
configuration is a field theory that lives on the $k$ D3-branes. At energies
below the Kaluza-Klein (KK) scale $m_{\rm KK}=\frac{1}{L}$ 
the effective theory is three-dimensional. In the presence of the $(1,N)$ fivebrane it
is known~\cite{Kitao:1998mf} that this theory is a $U(k)$ YM-CS theory at level
$N$ coupled to matter. The matter consists of three $\NN=1$ real scalar multiplets
in the adjoint of $U(k)$, associated with the directions $x^3$, $x^4$ and
$x^5$; we will denote them as $\Phi_3$, $\Phi_4$ and $\Phi_5$, respectively.
For generic angles $\psi$ and $\varphi$ these multiplets have independent masses 
and the field theory is  an $\NN=1$ YM-CSM theory.  

The three-dimensional gauge field $A$  is part of the $\NN=1$ vector multiplet which in addition
includes the gaugino Majorana fermion $\chi$. The scalar multiplets $\Phi_i$ include
a real scalar $\phi_i$ and a Majorana fermion $\psi_i$ $(i=3,4,5)$. The low-energy
effective action of these fields will be discussed in more detail in a moment.

The $\NN=1$ supersymmetry is enhanced to $\NN=2$ or $\NN=3$ for special
values of the angles $\psi$, $\varphi$. Enhancement to $\NN=2$ occurs when
$\psi=-\varphi$. The special case $\psi=\frac{\pi}{2}$ was the main focus of the
recent work~\cite{Giveon:2008zn} that formulated a Seiberg-like duality for
$\NN=2$ CSM theories. An extra set of $N_f$ D5-branes oriented along the
directions $(012789)$ was also present in that setup. Similar D5-branes and
their implications for $\NN=1$ CSM dynamics will be discussed in subsection
\ref{flavor}.

Further enhancement of the supersymmetry to $\NN=3$ occurs when all three
angles are correlated: $\psi=-\varphi=\theta$. A similar setup, with the $x^6$ direction compactified, was crucial in the
recent discussion of low-energy descriptions of the M2-brane worldvolume dynamics
based on CSM theories~\cite{Aharony:2008ug,Aharony:2008gk}.

One of the early motivations for studying the generic $\NN=1$ setup in fig.~\ref{sec2:orientN1} 
was to formulate the conditions for spontaneous breaking
of supersymmetry in Chern--Simons theories as a consequence of brane
dynamics~\cite{Witten:1999ds,Bergman:1999na,Ohta:1999iv}. These conditions
will play an important role in the next section.

\subsection{Getting the AV field theory}
\label{subsec:AV}

\begin{figure}[t!]
\centering
\includegraphics[height=5cm]{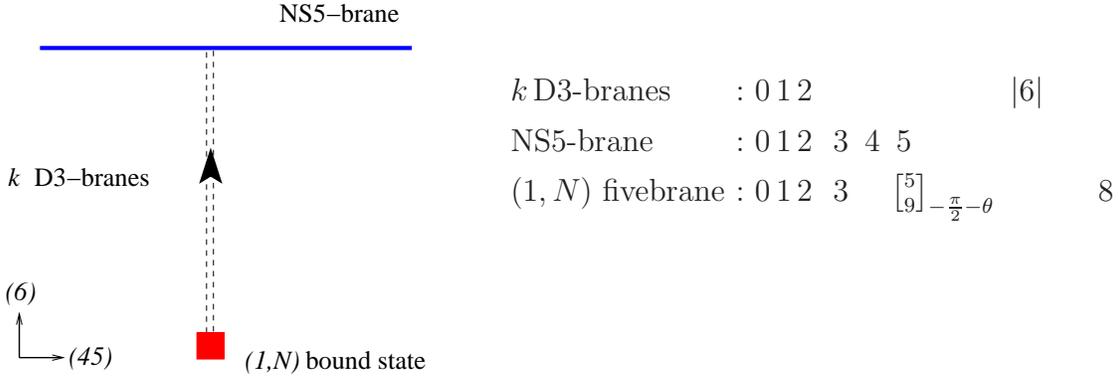}
\raisebox{3cm}{$\qquad \begin{array}{lclllll}
k\, \text{D3-branes} & : &0\, 1\, 2\, &&&&\, |6|\\
\text{NS5-brane} & : & 0\, 1\, 2& \, 3&\, 4&\, 5&  \\
(1,N) \ \text{fivebrane} &  : & 0\, 1\, 2 &\, 3 && \oao{5}{9}_{-\frac{\pi}{2}-\theta}& \qquad \quad 8   
\end{array}$}\bf
\caption{\it The special configuration that realizes the Acharya--Vafa CS theory.
Compared to the generic configuration in fig.\ \ref{sec2:orientN1} there is now
an extra common direction $(x^3)$ between the fivebranes.}
\label{AVelectricFig}
\end{figure}

The AV theory is a $U(k)$ $\NN=1$ CS theory at level $N$ coupled to a classically 
massless $\NN=1$ adjoint scalar multiplet. To recover this theory from the brane configuration in
fig.~\ref{sec2:orientN1} we must tune the angles $\psi$, $\varphi$ in such a way
that one of the scalar multiplets $\Phi_i$ becomes massless and the remaining two
extremely massive. This can be achieved, for example, by setting
\beq
\label{AVaa}
\psi=\frac{\pi}{2}~, ~~ \varphi=-\frac{\pi}{2}-\theta~,
\eeq
giving rise to a brane configuration with branes oriented as summarized in
fig.~\ref{AVelectricFig}.
$\Phi_3$ is now a massless scalar multiplet, $\Phi_4$ is infinitely massive and
$\Phi_5$ is massive with a mass $m_5$ that will be discussed in a moment.

Notice that by setting $N=0$ we replace the fivebrane bound state with an
NS5-brane which, according to equation \eqref{sec2:thetavalue},
is oriented along the directions (012389). This configuration preserves
four supersymmetries and gives rise to the three-dimensional $\NN=2$ SYM theory
on the D3-branes. With a non-zero value of $N$ we expect to 
have an extra $\NN=1$ Chern--Simons interaction at level $N$.
As reviewed in the introduction, this is the theory that according to~\cite{Acharya:2001dz}
describes the infrared dynamics on the domain walls of $\NN=1$ SYM in
four dimensions.

Before establishing this fact we have to tie a loose end. The low-energy field
theory on the D3-branes includes, for non-zero $N$, the extra massive scalar
multiplet $\Phi_5$. The mass of this multiplet is \cite{Barbon:1997zu,Ohta:1999iv}
\begin{equation}
\label{AVab}
m_5 = \frac{\cot\, \theta}{L}
=\frac{m_\textsc{kk}}{g_s N}
=\left( \frac{m_\textsc{kk}}{m_\textsc{cs}} \right)^2 \, m_\textsc{cs}
\, ,
\end{equation}
where $m_{\textsc{cs}}$ is the Chern--Simons induced mass of the gauge field
\beq
\label{AVad}
m_\textsc{cs}=g^2_{\textsc{YM}} N~,
\eeq
and $g_{\textsc{YM}}$ the dimensionful three-dimensional Yang-Mills coupling given by
the equation
\beq
\label{AVac}
\frac{1}{g^2_\textsc{ym}}=\frac{L}{g_s}
~.
\eeq
In the perturbative string regime of interest,
\beq
\label{AVae}
\frac{m_\textsc{kk}}{m_\textsc{cs}}=\frac{1}{g_s N} \gg 1~,
\eeq
the KK modes are very heavy and can be ignored. Then the hierarchy of scales
\begin{equation}
m_\textsc{cs} \ll m_\textsc{kk} \ll m_5
\label{sec2:masshiearch}
\end{equation}
guarantees that the multiplet $\Phi_5$ can be safely integrated out.

To obtain the precise Lagrangian of the low-energy theory on the brane setup of
fig.~\ref{AVelectricFig}, we can start from the $U(k)$ $\NN=3$ CS theory at level
$N$, whose Lagrangian is completely fixed by supersymmetry. 
It describes the low-energy dynamics of the brane configuration with
$\psi=-\varphi=\theta$. In components, the action of this theory 
reads~\cite{Kao:1992ig,Bergman:1999na}
\begin{align}
\SS =& \SS_\textsc{ym} + \SS_\textsc{cs}
\nonumber\\
= &\frac{1}{4g_\textsc{ym}^2\!\!} \int \text{d}^3x~\text{Tr} \Big\{\!
-F^2 \!+ \sum_{i=3}^{5} \left[ (D \phi_i)^2 + C_i^2
+ i \bar \psi_i  D\!\!\!\! \slash\, \psi_i\right]
+ i \bar{\chi}  D\!\!\!\! \slash\, \chi +i \bar{\chi}\nonumber \\&+
[\psi_3,\phi_3] + i \bar \chi [\psi_5,\phi_5]
 + i \bar \psi_3 [\chi,\phi_3]
+i \bar \psi_3 [\psi_5,\phi_4] + i \bar \psi_5 [\chi,\phi_5]\nonumber\\
& 
+  i \bar \psi_5 [\psi_3,\phi_4]-2i\bar{\chi}[\psi_4,\phi_4]
-2i\bar{\psi}_3[\psi_4,\phi_5]
+\tfrac{1}{2} \sum_{i<j}[\phi_i,\phi_j]^2\Big\}
\nonumber\\
& + \frac{N}{4\pi}\int  \tr \left(  A\wedge \text{d}A + \tfrac{2}{3} A^3\right)
\nonumber\\ 
&+\frac{N}{4\pi} \int \text{d}^3x \left \{-\bar{\chi} \chi
+ \sum_{i=3}^5 \left[(-)^i \bar{\psi_i} \psi_i + 2 \phi_i C_i\right]
+ \tfrac{1}{3} \sum_{ijk} \epsilon_{ijk} \phi_i[\phi_j,\phi_k]
\right\}~.
\label{sec2:CSlag}
\end{align}
The $C_i$ $(i=3,4,5)$ are auxiliary scalars in the $\Phi_i$ multiplets.

All the fields in the above action have the same mass whose value is fixed
by the $\mathcal{N}=3$ CS term in the last line of~\eqref{sec2:CSlag}.
Configurations with less supersymmetry can be obtained by changing the 
mass of each of the $\NN=1$ scalar multiplets separately. In particular, by 
tuning the bare mass of $\Phi_3, \Phi_4$ and $\Phi_5$ to zero, infinity and $m_5$,
respectively, integrating out $\Phi_4$ and $\Phi_5$ and renaming $\phi_3=\phi$,
$\psi_3=\psi$, we obtain the AV action, as described in the introduction,
\bea
\label{sec2:N1CSlag}
\SS_\textsc{AV}  
&=& \frac{1}{4g_\textsc{ym}^2\!\!} \int \text{d}^3 x ~\text{Tr} \left( \!
-F^2 \!+ (D \phi)^2 + i \bar \psi  D\!\!\!\! \slash\, \psi
+ i \bar{\chi}  D\!\!\!\! \slash\, \chi +2i\bar{\chi}[\phi,\psi] \right)
\nonumber\\
&&+ \frac{N}{4\pi} \int \text{Tr}  \left( A\wedge \text{d}A + \tfrac{2}{3} A^3\right)
-\frac{N}{4\pi} \int \text{d}^3x\,  \bar{\chi} \chi
~. 
\eea

\subsection{Comments on the dynamics of the AV theory}
\label{AVdynamics}

At low energies (below $m_\textsc{cs}$) the standard kinetic term
of the gauge field  and the kinetic term of the gaugino $\chi$ can
be dropped and the YM-CS action for the AV theory \eqref{sec2:N1CSlag}
becomes the action of $\NN=1$ CS theory coupled to a massless $\NN=1$
scalar multiplet,
\bea
\label{AVdynaa}
\SS_\textsc{AV-CSM}&=& \frac{N}{4\pi} \int \text{Tr}  \left( A\wedge \text{d}A + \tfrac{2}{3} A^3\right)
-\frac{N}{4\pi} \int \text{d}^3x\,  \bar{\chi} \chi
\nonumber\\
&&+\frac{1}{4g_\textsc{ym}^2\!\!} \int \text{d}^3 x ~\text{Tr} \left\{ \!
(D \phi)^2 + i \bar \psi  D\!\!\!\! \slash\, \psi
+2i\bar{\chi}[\phi,\psi] \right\}
~.
\eea
The massive gaugino can be integrated out to obtain
the classically marginal quartic interaction of the form $[\phi,\psi][\phi,\bar \psi]$.

$\NN=1$ supersymmetry is not enough to guarantee the absence of quantum
corrections to this action. In a similar situation with $\NN=2$ supersymmetry,
$e.g.$ the situation of $\NN=2$ CS theory coupled to an $\NN=2$ chiral multiplet
in the adjoint without superpotential interactions, it is known \cite{Gaiotto:2007qi}
that quantum effects do not generate any relevant interactions and hence the
theory is an exact CFT. With just $\NN=1$ supersymmetry relevant interactions
can and will be generated.

Indeed, one can show explicitly in the AV theory \cite{Armoni:2005sp,Armoni:2006ee}
that $\phi$ is not a true modulus and that quantum corrections lift the classical
moduli space parametrized by the vacuum expectation values (VEVs) of gauge-invariant 
polynomials in $\phi$. The lifting is a $1/N$ effect. The reason is that in the 
large $N$ limit the $k$-wall becomes a collection of $k$ non-interacting fundamental 
domain walls that can be separated freely. This is in agreement with our brane picture 
where in the limit $N\rightarrow \infty$, $\theta \rightarrow \pi /2$ and the theory acquires a 
quantum moduli space.

One can split the scalar multiplet as $\Phi=\Phi_0 + \hat{\Phi}$ according to the decomposition 
$u(k) \simeq u(1) \oplus su(k)$. Performing the calculation of a two-loop Coleman-Weinberg 
effective potential in the Coulomb branch of the $U(2)$ AV theory, one finds a potential of the 
form \cite{Armoni:2005sp}
\beq
V(u) \sim  {1 \over N} {u  \over 1+u} ~,
\label{potential}
\eeq
where $u \equiv \langle \text{Tr}(\hat{\phi}^{\, 2}) \rangle /m_{\textsc{cs}}^2$. 
This perturbative result captures the leading $1/N$ effects. Higher order
corrections are expected to modify the potential \eqref{potential}.
In general, an attractive potential is generated near the origin for $u$ 
but no potential is generated for the overall `center-of-mass' VEV $\langle \phi_0 \rangle$. 
We will re-encounter these quantum effects in the next section where we discuss the
brane dynamics in the type IIB string theory setup of fig.\ \ref{AVelectricFig}.

We observe that the leading term of the potential \eqref{potential} is a quadratic term
with mass
\be
\label{massloop}
m _{\textsc{loop}}=\frac{m_{\textsc{cs}}}{N}~,
\ee
which is parametrically smaller than the CS mass $m_{\textsc{cs}}$ in the large-$N$
limit. Hence, at energies below $m_{\textsc{loop}}$, the $U(k)$ CSM theory
becomes a topological field theory -- the bosonic CS theory -- with an additional
$decoupled$ free massless real scalar field $\phi_0$ and its superpartner. With the exception of
the decoupled massless scalar multiplet we would have obtained the same infrared
dynamics for any of the low-energy theories that live on the D3-branes of the general
setup in fig.\ \ref{sec2:orientN1}, where a mass is present for the scalar multiplets
already in the tree-level Lagrangian.

\subsection{T-duality and the AV string theory setup}
\label{Tduality}

The brane system in type IIB string theory that appears in fig.\ \ref{AVelectricFig} is
related to the Acharya--Vafa setup in type IIA string theory, but exhibits some 
differences. To see the relation, one may compactify $x^7$ on a circle and T-dualize 
along this direction. Before analyzing this transformation it will be useful to recall 
the setup of Acharya and Vafa in type IIA string theory.

The starting point is type IIA string theory on $\R^{3,1}$ times the deformed conifold.
Wrapping $N$ D6-branes around the non-vanishing three-cycle 
one obtains at low energies on the D6-branes four-dimensional $\NN=1$ SYM. It has 
been argued~\cite{Vafa:2000wi} that there is a large-$N$ holographic
description of this theory which involves a geometric transition from the deformed
conifold to a resolved conifold. The D6-branes disappear in the resolved conifold 
and get replaced by their RR flux going through the non-vanishing two-cycle of the 
blown-up singularity. A string propagating in this background can be interpreted as 
the QCD-string of $\NN=1$ SYM. D4-branes wrapping the non-vanishing two-cycle 
of the resolved conifold are interpreted as domain walls in the $\NN=1$ SYM theory.

Now let us return to our setup of fivebranes. Consider first the case with $N=0$. Then 
the setup in fig.~\ref{AVelectricFig} consists of two NS5-branes, respectively along 
$(012345)$ and $(012389)$, and $k$ D3-branes along $(012|6|)$. T-dualizing along 
$x^7$ transforms the system of two fivebranes into the resolved conifold, whose blow-up 
parameter is controlled by $L$~\cite{Aganagic:1999fe,Ohta:1999we}. This type IIA setup 
is identical to the setup of ref.~\cite{Vafa:2000wi} after the large $N$ transition.
The D3-branes are mapped to wrapped D4-branes and there is no RR flux through 
the two-cycle.\footnote{The T-dual of Vafa's setup before the transition is given by a 
`brane box' model~\cite{Hanany:1998it} with $N$ D5-branes filling a disc bounded 
by an NS5-brane.} 

Consider now what happens when we replace the second NS5-brane by the 
$(1,N)$  bound state that has a modified orientation along 
$(01238\left[ {5 \atop 9}\right]_{-\frac{\pi}{2}-\theta})$. 
After the same T-duality along $x^7$ one gets a resolved conifold with RR flux and 
D4-branes around the two-cycle. More specifically, the NS5-brane turns into a 
Kaluza-Klein (KK) monopole stretched in the directions (012345), whose charge 
is associated with the T-dual of $x^7$, and the $(1,N)$ bound state turns into a 
U-dual of the KK dyon of \cite{Sen:1997zb}, namely, an RR flux on a (differently oriented) 
KK monopole. Combining the KK monopole and dyon gives rise to a resolved conifold 
with $N$ units of RR flux. However, the RR flux is oriented in different directions compared 
to the setup of ref.\ \cite{Vafa:2000wi}. 

Despite the differences between the T-dual of our setup and the setup of~\cite{Vafa:2000wi}, 
we have seen that the IR theory on the $k$ D-branes is the same in both cases and 
describes the low-energy dynamics of the domain wall that interpolates between the 
$\ell$-th and the $(\ell+k)$-th vacuum in $\NN=1$ SYM.

\section{D-brane Dynamics and Domain Walls in 4D SYM}
\label{sec:HWSYM}

Having established a relation between the brane configuration in fig.\ \ref{AVelectricFig}
and the four-dimensional $\NN=1$ SYM theory we now proceed to explore how
known facts in one theory map to known facts in the other. We will find that
some properties are easy to establish in one formulation and difficult in the
other making this comparison a fruitful exercise for both theories.

\subsection{Pseudo-moduli, the s-rule and brane creation}
\label{srule}

We have argued that the $\NN=1$ CS theory on $k$ suspended D3-branes
is coupled to an $\NN=1$ chiral multiplet $\Phi$ in the adjoint of the $U(k)$
gauge group. The vacuum expectation values of gauge-invariant polynomials 
in the scalar field $\phi$ parametrize the positions of the D3-branes 
in the transverse direction $x^3$. Classically, there is no potential for 
them and the three-branes can move apart without any cost of energy.

In brane setups with more supersymmetry ($\NN=2$ and higher) similar
(complex) moduli exist both classically and quantum mechanically. With
just $\NN=1$ supersymmetry, however, quantum corrections  lift the 
classical moduli space. As we reviewed in
subsection~\ref{AVdynamics}, in field theory these effects generate a
Coleman-Weinberg potential for the VEVs associated with the $SU(k)$  adjoint scalar 
$\hat{\phi}$ (see eq.~\eqref{potential} for the $k=2$ case, $i.e.$ for two domain walls) 
stabilizing them at the origin. The $U(1)$ part $\langle \phi_0 \rangle$ remains a 
modulus which is consistent with the expectation that we can arbitrarily place the 
center of mass of $k$ D3-branes at any point along the $x^3$ direction.

Since there are no tachyons along the classical moduli space a phase transition
is not anticipated as we go from the field theory regime to the perturbative
brane regime. Consequently, a similar stabilization of the pseudo-modulus
at the origin is expected also in the perturbative brane regime. We will present
an alternative indirect argument for this in a moment. For two or more D3-branes
stretching between the fivebranes in fig.~\ref{AVelectricFig} the stabilization of
the pseudo-modulus implies that while we can freely move their center of mass
along the $x^3$ direction, we cannot separate them without some cost of energy.
An attractive force between the D3-branes forms a bound state that behaves
as a single object.

The formation of D3-brane bound states in our setup, labelled by $k$,
matches nicely what is expected from domain walls in the $\NN=1$
SYM theory. The bound states have a non-trivial tension which is a certain
function of the parameters $k$ and $N$. The form of this function
will be discussed shortly. Now we want to discuss the precise range of
the rank $k$ of the gauge group.

Supersymmetry restricts the number $k$ of D3-branes that can be suspended
between the fivebranes in the brane setup of fig.~\ref{AVelectricFig}.
The standard $s$-rule of brane dynamics continues to hold in our case
and dictates that the configuration is supersymmetric if and only if $k\leqslant N$.
It is particularly interesting to pinpoint the ingredients that conspire to make the
$s$-rule work in our setup.

A standard argument for the validity of the $s$-rule is the following.
By moving the $(1,N)$ bound state along the $x^6$ direction past the
NS5-brane $k$ D3-branes are carried along and become anti-D3-branes.
During the crossing of the fivebranes $N$ D3-branes are created via the
brane creation effect~\cite{Hanany:1996ie}. The brane creation effect
ensures that the dynamics is smooth during the crossing and that the
amount of supersymmetry of the original configuration is preserved in
the final configuration. Hence, in accordance with the $s$-rule of the
original setup, for $k\leqslant N$ the annihilation of $k$ brane/anti-brane pairs
leaves behind a supersymmetric configuration of $N-k$ suspended D3-branes.
In the opposite regime ($k>N$) the annihilation of $N$ brane/anti-brane pairs 
leaves behind a non-supersymmetric configuration of $k-N$ suspended
anti-D3-branes.

In the absence of the attractive potential between the D3-branes
a contradiction with the $s$-rule would have been obtained. We would
have been able to freely separate the D3-branes along the $x^3$ direction
to obtain a supersymmetric configuration for any $k$. Hence, the validity
of the $s$-rule requires the presence of the attractive potential
between the D3-branes in the brane regime and confirms our expectations
from field theory~\cite{Armoni:2005sp,Armoni:2006ee}.

The Witten index corroborates this picture. The validity of the $s$-rule
requires that the Witten index is non-zero for $k\leqslant N$ in field theory and zero
in the opposite regime. Indeed, for the AV CS theory the Witten index was
computed in \cite{Acharya:2001dz} and was found to be proportional to
$\frac{1}{(N-k)!}$. At the same time, the independence of the Witten index
from the mass of $\hat{\Phi}$ \cite{Acharya:2001dz}
fits nicely with the fact that the standard $s$-rule holds for general
angles in the brane setup of figure \ref{sec2:orientN1}, where $\Phi$ is
massive. In the next subsection we will further show how branes provide
a natural geometric interpretation of the precise value of the Witten index
for arbitrary values of $k$.

Since there is an upper bound on the number $k$ of D3-branes
that can be stretched between fivebranes in our setup without breaking
supersymmetry we conclude that there are only $N$ distinct supersymmetric
D3-brane bound states for $k=1,2,\cdots,N$. Later we will see that the
$N$-th state with $k=N$ is equivalent to the pure vacuum $k=0$ state,
hence the number of distinct D3-brane bound states obtained in this way is actually
$N-1$. This number matches exactly the number of different domain walls
in the four-dimensional $\NN=1$ SYM  theory and is further evidence for the
validity of the above picture.

\subsection{Witten index and the degeneracy of domain walls}
\label{index}

As we reviewed in the introduction, the $k$-th domain wall in the $\NN=1$ SYM
theory has a degeneracy given by the index $I_{k,N}$, presented in eq.~\eqref{indexaa}.
In the CS theory that captures the low-energy dynamics of the $k$-th domain wall
this index counts the number of supersymmetric vacua. We can ask whether this
degeneracy is visible in the brane setup of fig.~\ref{AVelectricFig}.

At first sight we seem to get a different answer. $k$ D3-branes stretch between
an NS5-brane and a $(1,N)$ fivebrane in an apparently unique way. 
A similar mismatch between~\eqref{indexaa} and the
counting of BPS branes in string theory was also observed in the type IIA context of
ref.~\cite{Acharya:2001dz} and earlier in the context of MQCD in \cite{Volovich:1998bv}.
Ref.~\cite{Acharya:2001dz} pointed out that this issue is related to the global boundary
conditions on the domain walls. Counting vacua in Minkowski space is different
from counting vacua in a toroidally compactified theory. The computation of the Witten
index in the toroidally compactified AV CSM theory gives the expected 
answer~\eqref{indexaa}~\cite{Acharya:2001dz}. Hence, we can now refine the above
question. Can we reproduce the index \eqref{indexaa} in the brane setup
of fig.\ \ref{AVelectricFig} after compactifying the worldvolume of the D3-branes?

The answer, which has a simple geometric interpretation, lies already in a work 
by Ohta \cite{Ohta:1999iv}. We will review the argument here with the appropriate
modifications. Consider compactifying one of the worldvolume directions of the D3-branes,
say the direction $x^2$. T-dualizing the setup along $x^2$, adding the M-theory circle $x^{10}$
and lifting to M-theory we obtain a configuration with M2-branes along $(01|6|)$ stretched
between an M5-brane along $(012345)$ and an M5-brane along
$(0138\left[ {5 \atop 9}\right]_{-\frac{\pi}{2}-\theta})$ wrapping the cycle
$\alpha_2+N\alpha_{10}$, where $\alpha_2$ is the cycle associated with the direction
$x^2$ and $\alpha_{10}$ the cycle associated with the direction $x^{10}$.
The M5-branes intersect $N$ times on the $(2,10)$ torus. An M2-brane stretching
between them without breaking the supersymmetry is necessarily attached on each
of these $N$ intersection points. According to the $s$-rule no more than one M2-brane
can be attached to the same intersection point if we require supersymmetry. Hence, the
counting of possible supersymmetric configurations of stretched M2-branes boils
down to a counting of $k$ M2-branes distributed along the $N$ M5 intersection
points without violating the $s$-rule. For $k\leqslant N$ the answer is trivially 
given by eq.~\eqref{indexaa}.
For $k>N$ the $s$-rule is necessarily violated and no supersymmetric vacuum
exists. This picture provides a simple, intuitive understanding of the Witten 
index~\eqref{indexaa} in string/M theory.

\subsection{Tension formula}
\label{tension}

Another characteristic feature of the BPS domain walls in $\NN=1$ SYM theory
is their tension, given by the sine formula \cite{Dvali:1996xe},
\beq
\label{tensionaa}
T_k=\frac{N^2\Lambda^3}{4\pi^2}\sin\frac{\pi k}{N}~,
\eeq
for the $k$-th domain wall, in terms of the dynamically generated QCD scale $\Lambda$.

Because of the attractive potential between the D3-branes, the $k$-th domain wall maps
to a bound state of $k$ D3-branes in our setup. The attractive force is weak for
$\frac{k}{N}\ll 1$ and to leading order in $\frac{k}{N}$ the tension of $k$ suspended
D3-branes is
\beq
\label{tensionab}
T_k = k T_1  + \OO(N^{-1})
\sim \frac{L}{g_s \ell_s^4} k+\OO(N^{-1})
~.
\eeq
Identifying
\beq
\label{tensionac}
g_s \sim \frac{L}{N\ell_s}~, ~ ~ \ell_s\sim \frac{1}{\Lambda}~,
\eeq
we reproduce (up to a numerical coefficient) the leading order term of $T_k$ in
\eqref{tensionaa}. The identification \eqref{tensionac} assumes parameters typical
of large-$N$ string theory duals of a QFT (see $e.g.$ \cite{Vafa:2000wi,Armoni:2005sp}).
Hence, the string coupling $g_s$ is proportional to $\frac{1}{N}$, and the string scale 
$\ell_s$ is set by the QCD string tension. Notice that in this identification 
$m_\textsc{cs}\sim \Lambda$.

Reproducing the full $k/N$ dependence of $T_k$ requires a difficult non-perturbative
computation in CS theory. A perturbative treatment of the tension in the AV CS-YM
theory can be found in \cite{Armoni:2005sp,Armoni:2006ee}. From the point of view
of the type IIB setup in fig.\ \ref{AVelectricFig} an exact computation of $T_k$ requires
a detailed knowledge of the forces that form the D3-brane bound states, which is
currently lacking. It is interesting, however, that by turning the tables around and
viewing the D3-branes as domain walls in the four-dimensional $\NN=1$ SYM
theory we can determine exactly the $k$ D3-branes tension as in \eqref{tensionaa}.

We conclude by noticing the identity
\beq
\label{tensionad}
T_{k}=T_{N-k}
~.
\eeq
As we will review in a moment this identity has a natural explanation in the $\NN=1$
SYM theory. We will be able to recover it independently in our brane setup together
with a statement of Seiberg duality for the $\NN=1$ AV CSM theory.

\subsection{Seiberg duality}
\label{duality}

We are now coming to a different aspect of the dynamics of the $\NN=1$ AV CSM
theory: a Seiberg-like duality that relates the $U(k)$ description to a $U(N-k)$ one.

Let us first review briefly a related example with $\NN=2$ supersymmetry
that appeared in~\cite{Niarchos:2008jb,Niarchos:2009aa}. It involves the $U(k)$
$\NN=2$ CS theory at level $N$ coupled to an $\NN=2$ chiral multiplet $X$ in the
adjoint representation of the gauge group. To formulate Seiberg duality in this case
a tree-level superpotential is necessary,
\beq
\label{dualityaa}
W_{n+1}=g_{n} \tr X^{n+1}~, ~~ n\geq 1
~.
\eeq
The dual is a  level $N$ $U(nN-k)$ CS theory with an adjoint and a superpotential \eqref{dualityaa} .
In the special case of
$n=1$ the superfield $X$ is massive and can be integrated out to recover the
$\NN=2$ CS theory which is a topological field theory. By a standard argument
that will be reviewed in a moment, duality in CS theory reduces, in this case, to
level-rank duality in an $SU(N)$ WZW model.

The situation in the AV theory is similar, but instead of an $\NN=2$ CS theory
coupled to an $\NN=2$ chiral multiplet in the adjoint we have an
$\NN=1$ CS theory coupled to an $\NN=1$ adjoint multiplet. A tree-level
potential is absent, but one is generated at the quantum level. In that sense,
the AV theory is similar to the $n=1$ special case of \eqref{dualityaa}.
A Seiberg duality between the $U(k)$ AV theory and the $U(N-k)$ AV theory 
(both at level $N$) is anticipated.

In the deep IR (below the energy scale $m_{\textsc{loop}}$ set by the loop corrections
-- see eq.\ \eqref{massloop}) one is left with the bosonic $U(k)$ CS theory plus
a decoupled free massless real scalar field and  a fermion. Integrating out the massive
fermions gives rise to a shift of the level from $N$ to $N-k$~\cite{Kao:1995gf} (the contribution
of the gauge field is not included here). Hence, the IR theory includes a level
$N-k$ $SU(k)$ pure CS theory. In the infrared, the rank of this dual theory is 
similarly shifted from $N$ to $k$, giving rise at low energies to an $SU(N-k)$ pure 
CS theory at level $k$. The $SU(k)$ level $N-k$ and the $SU(N-k)$ level $k$ theories 
are equivalent by the CS-WZW correspondence of~\cite{Witten:1988hf} and level-rank 
duality in $SU(N)$ WZW models. In this way, we recover a duality between
the $U(k)$ and $U(N-k)$ AV theories in the infrared.

Further aspects of this duality can be deduced by reformulating it in the context 
of string theory and in the context of the four-dimensional $\NN=1$ SYM theory, 
as we discuss next.

\subsubsection*{Seiberg duality from brane dynamics}
\label{dualbranes}

As we mentioned above, in the type IIB setup of fig.~\ref{AVelectricFig} no
phase transitions are anticipated as we change continuously the
separation $L$ of the NS5-brane and the $(1,N)$ bound state along the
$x^6$ direction. Even when we pass the $(1,N)$ fivebrane through the
NS5-brane the process is smooth and the IR dynamics on the worldvolume
of the D3-brane remains invariant.

\begin{figure}[t!]
\centering
\includegraphics[height=5cm]{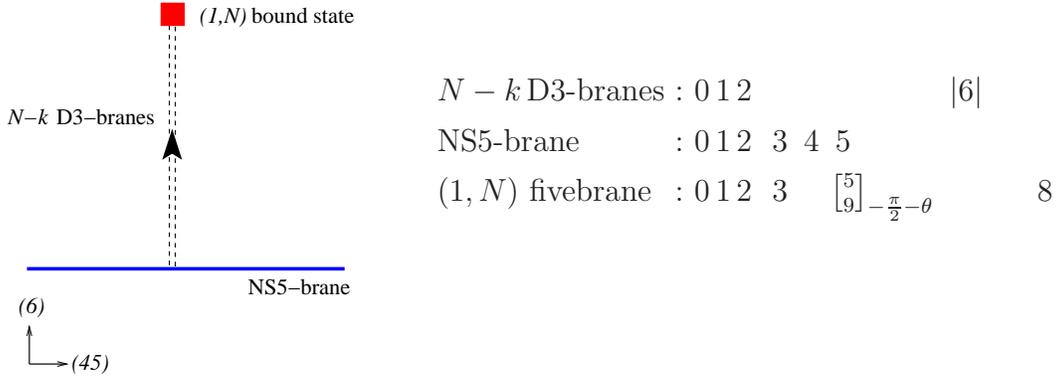}
\raisebox{3cm}{$\qquad  \begin{array}{lclllll}
N-k\, \text{D3-branes} & : &0\, 1\, 2\, &&&&\, |6|\\
\text{NS5-brane} & : & 0\, 1\, 2& \, 3&\, 4&\, 5&  \\
(1,N) \ \text{fivebrane} &  : & 0\, 1\, 2 &\, 3 && \oao{5}{9}_{-\frac{\pi}{2}-\theta}& \qquad \quad 8   
\end{array}$}\bf
\caption{\it The brane setup that realizes the dual Acharya--Vafa CS theory.}
\label{AVmagneticFig}
\end{figure}

We have already observed that by passing the $(1,N)$ fivebrane through the
NS5-brane one is left with the configuration of fig.~\ref{AVmagneticFig} that involves
$N-k$ suspended D3-branes. The new configuration realizes a dual description
of the original $U(k)$ AV CS theory. Below the energy scale set by $m_\textsc{cs}$
this description is still an $\NN=1$ CSM theory at level $N$ coupled to a massless
adjoint multiplet with a dual gauge group $U(N-k)$. In the dual description the
supersymmetry conserving condition $k\leqslant N$ becomes a classical fact related
to the rank of the gauge group in the dual theory. This is typical in dualities of this kind.

\subsubsection*{Seiberg duality as a charge conjugation symmetry}
\label{dualconj}

The embedding of the AV theory into the four-dimensional $\NN=1$ SYM
theory provides yet another way to look at this duality.

From the $\NN=1$ SYM point of view the $U(k)$ AV CS theory at level $N$
describes the IR dynamics of the theory that lives on the $k$-th domain wall
which interpolates `clockwise' between the $j$-th and the $(j+k)$-th vacuum.
By charge conjugation symmetry this domain wall is equivalent to the $(N-k)$-th
anti-wall that interpolates `anti-clockwise'  between the $(j+N-k)$-th and the $j$-th
vacuum. The IR physics of the theory that lives on this anti-wall is captured by the
$U(N-k)$ dual AV CS theory. We observe that in this case Seiberg duality, which is
a hard non-perturbative statement in three-dimensions, becomes a statement
that follows directly from a simple symmetry, $i.e.$ charge conjugation symmetry,
in the `parent' four-dimensional gauge theory. It would be interesting to
know if there are other examples in field theory where Seiberg duality
can be derived in this way.

The duality between the $U(k)$ and the $U(N-k)$ $\NN=1$ AV theories
is a strong/weak coupling duality. The 't Hooft coupling of the $U(k)$ 
theory~(\ref{AVdynaa}) is $\lambda=\frac{k}{N}$ whereas that of the $U(N-k)$ 
theory is
\beq
\label{dualbranesaa}
\widetilde\lambda=\frac{N-k}{N}=1-\lambda
~.
\eeq
The strongly coupled point at $k=\frac{N}{2}$ (for $N$ even) is self-dual under
the duality. It would be interesting to know if the theory enjoys special properties
at this point.

At low energies below $m_\textsc{loop}$, where the theory is topological,
the duality is, as we mentioned, a consequence of level-rank duality in WZW models.
The observables of the bosonic CS theory are Wilson loop operators \cite{Witten:1988hf}.
Charge conjugation symmetry in the context of $\NN=1$ SYM theory predicts that
the expectation values of these Wilson loop operators are invariant under the
replacement $k \rightarrow N-k$. This is consistent with level-rank duality.

In addition, the D-brane and domain wall perspectives from string theory and
the $\NN=1$ SYM theory, respectively, suggest that the duality extends beyond
the topological data of the IR theory below $m_\textsc{loop}$. Note that the
tension of the domain walls can be calculated using the AV theory.
Specifically, when $N$ is large $T_k = kT_1 + V$, where $V$ is the binding energy.
The binding energy is calculated by a Coleman-Weinberg potential. The tension, which
is {\it not} a topological datum, admits $T_k = T_{N-k}$, namely a Seiberg dual relation.
Therefore, it suggests that the equivalence goes beyond the topological data and is
valid throughout the whole RG flow from the scale  $m_\textsc{cs}$ to the far IR.

\section{Generalizations}
\label{sec:general}

The type IIB brane construction suggests a number of interesting
generalizations of the $\NN=1$ AV CSM theory. Two of them will
be discussed briefly in what follows. In the first example we consider
extra matter in the fundamental representation of the gauge group.
The second example, which includes additional matter in the adjoint,
is a particularly interesting case where one can argue, in a certain
regime of parameters, that the infrared theory is an interacting conformal
field theory, instead of a topological field theory.

\subsection{Adding flavor}
\label{flavor}

\begin{figure}[t!]
\centering
\includegraphics[height=7cm]{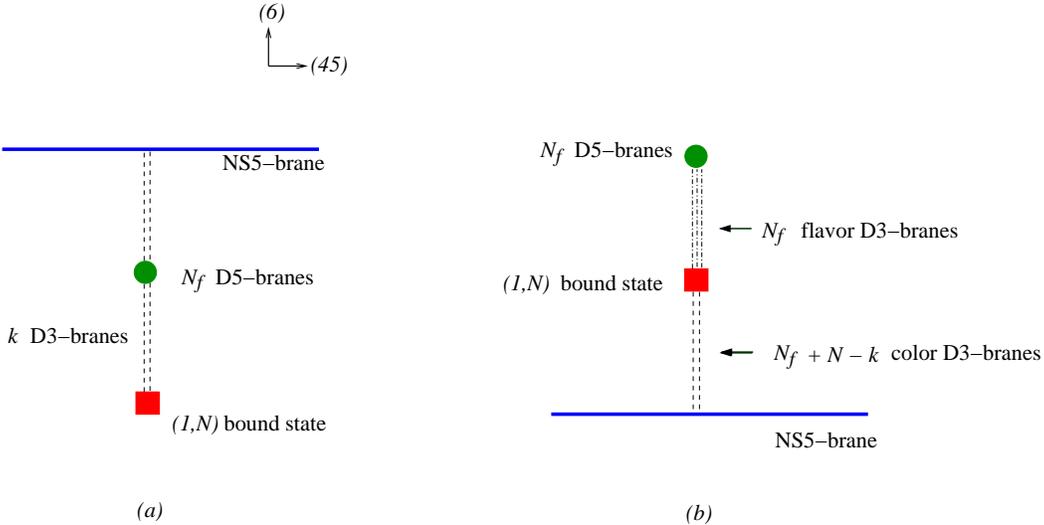}\bf
\caption{\it Fig.\ $(a)$ depicts the brane setup that realizes a
flavored version of the $U(k)$ Acharya--Vafa CS theory. The setup
in fig.\ $(b)$ realizes its $U(N_f+N-k)$ dual. The orientation of the D3-branes,
the NS5-brane and the fivebrane bound state is the same as that in fig.\
\ref{AVelectricFig}. The D5-branes are oriented along the directions
(012789).}
\label{AVflavor}
\end{figure}

Adding $N_f$ D5-branes, oriented along the directions (012789),
in the brane setup of fig.~\ref{AVelectricFig} we obtain the configuration
of fig.~\ref{AVflavor}$(a)$. In this configuration the low-energy
theory on the D3-branes becomes a $U(k)$ $\NN=1$ CS-YM theory at
level $N$ coupled to a massless $\NN=1$ scalar multiplet in the adjoint
and $N_f$ pairs of $\NN=2$ chiral multiplets $Q^i$, $\widetilde Q_i$
$(i=1,\cdots, N_f)$ in the fundamental and anti-fundamental
representations of the gauge group.\footnote{One can also consider adding
$N_f$ D5-branes in the more general brane setup of fig.~\ref{sec2:orientN1}.
In that case, the $\NN=1$ scalar multiplet is massive
and comes along with two additional massive $\NN=1$ scalar multiplets.
The resulting setup, which will not be discussed explicitly here, bears
many similarities with the $\NN=2, 3$ setups analyzed in~\cite{Giveon:2008zn}.}

The Lagrangian that describes the low-energy dynamics of D3-branes in the
setup that appears in fig.~\ref{AVflavor}$(a)$ includes:
$(i)$ the $\NN=1$ AV CS-YM Lagrangian \eqref{avac}-\eqref{avafull} for the
$\NN=1$ vector multiplet and the adjoint superfield $\Phi_3$, $(ii)$ the
standard $\NN=2$ kinetic terms for the quark multiplets $Q^i$, $\widetilde Q_i$,
and $(iii)$ a tree-level superpotential coupling between the multiplets $\Phi_3$
and $Q^i$, $\widetilde Q_i$,
\beq
\label{flavoraa}
\int \text{d}^2\theta ~ Q^i \Phi_3 \widetilde Q_i+c.c.~,
\eeq
written here in the $\NN=2$ formalism with $\Phi_3$ regarded as the
real part of an $\NN=2$ chiral superfield.

The infrared dynamics of this theory is controlled by several factors.
The cubic coupling \eqref{flavoraa} is classically relevant and affects the
RG flow, but there are additional interactions generated by loop effects.
In particular, there is no symmetry preventing the generation of masses
for the matter fields. At weak coupling, one expects that the quantum
generated mass terms take over. In the deep infrared this leads again to a
description in terms of the topological CS theory. It is unclear if a non-trivial
interacting fixed point can arise at large values of the effective coupling
$k/N$.

From the string theory embedding we can read off immediately the
following properties. First, the $s$-rule of brane dynamics
dictates that the theory exhibits spontaneous breaking of supersymmetry when
\beq
\label{flavorab}
k > N_f+N
~.
\eeq
Second, by passing the $N_f$ D5-branes and the $(1,N)$ bound state
through the NS5-brane along  $x^6$  we obtain the configuration
in fig.~\ref{AVflavor}(b) that realizes a dual $U(N_f+N-k)$ $\NN=1$ CSM theory
at level $N$ coupled to the following matter multiplets: $(i)$ an $\NN=1$ scalar multiplet 
$\widetilde\Phi_3$, $(ii)$ $N_f$ pairs of dual $\NN=2$ quark multiplets $q_i$,
$\tilde q^i$, and $(iii)$ a set of gauge-singlet dual  meson $\NN=1$
scalar multiplets $M_i^j$ $(i,j=1,\cdots,N_f)$. The scalar component of the
$\widetilde\Phi_3$ multiplet describes the fluctuations of the $N-k$ color D3-branes
in the $x^3$ direction. The scalar components of the $M_i^j$ multiplets describe
the fluctuations of the $N_f$ flavor D3-branes in the $x^8$ direction which is
common to the D5 and $(1,N)$ branes. The dual quarks arise from the open
strings stretching between the color and flavor D3-branes.

The dual theory possesses the tree-level superpotential interaction
\beq
\label{flavorac}
\int \text{d}^2 \theta \left( \tilde q^i \widetilde\Phi_3 q_i
+M_i^j \tilde q^i q_j\right)+c.c.
~.
\eeq
A short explanation of this interaction proceeds in the following way.
The flavor D3-branes are stuck at $x^3=0$. By moving the color D3-branes
in the $x^3$ direction the dual quarks become massive, a fact which is captured
by the first cubic interaction in the above superpotential. Similarly, the color
D3-branes are stuck at $x^8=0$, whereas the flavor D3-branes can move in
the $x^8$ direction making again the dual quarks massive. This fact is captured
by the second cubic interaction in \eqref{flavorac}.

\subsection{Adding an adjoint superfield with a tree-level superpotential}
\label{adjoint}

Another interesting generalization involves taking a general number $n$ of
NS5-branes in the brane configuration of fig.\ \ref{AVelectricFig}. For simplicity,
we set the number of D5-branes $N_f$ to zero, but analogous statements
can be made in the more general case. Then, the low-energy theory
on the $k$ suspended D3-branes is a $U(k)$ $\NN=1$ CS theory at level $N$
coupled to two adjoint $\NN=1$ superfields $\Phi$ and $X$. Once again, $\Phi$
is a massless multiplet whose lowest scalar component describes the fluctuations
of the D3-branes in the $x^3$ direction. $X$ is a multiplet whose lowest scalar
component describes the fluctuations of the D3-branes in the $x^8$ direction.
Since $x^8$ is not a common direction of the NS5 and $(1,N)$ branes the motion
of the D3-branes along $x^8$ is not free. In the low-energy field theory on the
D3-branes this effect is captured by a tree-level $\NN=1$ superpotential $W(X)$
of degree $n+1$.

The precise form of $W$ is closely related to the one-dimensional modulus
of the brane setup that controls the $x^8$ position of the $n$ NS5-branes.
Placing the NS5-branes at $n$ different points $x^8_j$, $j=1,\cdots,n$, forces
the $k$ D3-branes to break up into $n$ groups of $r_j$ D3-branes ending on the
$x_j^8$ positioned NS5-brane with
\beq
\label{adjointaa}
\sum_{j=1}^n r_j=k
~.
\eeq
{From} the D3-brane point of view $x_j^8$ are the real expectation values of the
diagonal matrix elements of the scalar component of the superfield $X$. In field
theory these vacua are accounted for by the $\NN=1$ superpotential
\beq
\label{adjointab}
W(X)=\sum_{j=1}^n \frac{s_j}{n+1-j}X^{n+1-j}
~.
\eeq
For generic coefficient $\{ s_j \}$ the superpotential has $n$ distinct minima
$\{x_j^8 \}$ related to $\{ s_j \}$ via the relation
\beq
\label{adjointac}
W'(x)=\sum_{j=0}^n s_j x^{n-j}=s_0 \prod_{j=1}^n (x-x_j^8)
~.
\eeq
The integers $(r_1,\cdots,r_n)$ label the number of the eigenvalues of the
$N_c\times N_c$ matrix $X$ residing in the $j$-th minimum (for $r_j$) of
the scalar potential $V=|W'(x)|^2$. When all the expectation values $x_j^8$
are distinct the adjoint field is massive and the gauge group is Higgsed:
\beq
\label{adjointad}
U(k) \to U(r_1) \times \cdots \times U(r_n)
~.
\eeq
In this vacuum we recover $n$ decoupled copies of the $\NN=1$ AV CSM
theory at level $N$.

At the origin of the NS5-brane moduli space (all $x_j^8=0$) the tree-level
superpotential is $W(X) \sim \tr X^{n+1}$. At weak coupling, $k/N\ll 1$, this
is an irrelevant operator (for $n>3$) that does not affect the IR dynamics. We
will see in a moment that this is not true for sufficiently large coupling. At the same
time quantum corrections generate a potential for both $\Phi$ and $X$.

{From} the $s$-rule of brane dynamics we learn that this theory has a
supersymmetric vacuum if and only if
\beq
\label{adjointae}
k \leqslant n N
~.
\eeq
We also learn, by exchanging the fivebranes, that there is a Seiberg dual
description in terms of a $U(nN-k)$ $\NN=1$ CS theory at level $N$
coupled again to two $\NN=1$ multiplets $\widetilde \Phi, \widetilde X$ 
in the adjoint.

As in the closely related $\NN=2$ examples of \cite{Niarchos:2009aa}, the
$s$-rule and Seiberg duality reveal some of the non-trivial properties of this
theory. In particular, we learn that by increasing the large-$N$ coupling
$\lambda=k/N$ there is a point $\lambda^\textsc{susy}_{n+1}=n$ where
supersymmetry gets spontaneously broken in the presence of the tree-level
deformation $\tr X^{n+1}$. This implies that there is a critical coupling
$\lambda^*_{n+1}<n$ beyond which the operator $\tr X^{n+1}$ becomes
relevant and affects the infrared dynamics. The Seiberg dual theory, which
is weakly coupled when $k/N$ is close to $n$, implies similarly that there is
an upper value $\lambda^{**}_{n+1}$ for $k/N$ above which the IR theory is 
again unaffected by the $\tr X^{n+1}$ deformation.

The picture that seems to be emerging from this information is the following.
At weak coupling the dynamics of the $U(k)$ theory at level $N$ is controlled
by the loop-generated effects and is described at low energies by the pure
CS theory. Duality relates this topological theory to a strongly coupled $U(nN-k)$
theory with a relevant $\tr X^{n+1}$ deformation.

As we further increase the coupling we encounter a regime of parameters (in
the large-$N$ limit this regime is given by $k/N \in [\lambda^*_{n+1},\lambda^{**}_{n+1}]$),
where the theory flows in the infrared to an interacting fixed point. This conformal
window appears in a non-perturbative region of the theory where both the $U(k)$
and $U(nN-k)$ descriptions of the theory are simultaneously strongly coupled. Notice
that in postulating this conformal window we have assumed the inequality
$\lambda^*_{n+1}<\lambda^{**}_{n+1}$. If the opposite inequality were true, we would have
obtained an inconsistency. Inside the range $[\lambda^{**}_{n+1},\lambda^*_{n+1}]$ the
loop-generated mass terms would dominate the IR dynamics and both the
$U(k)$ and $U(nN-k)$ theories would be described at low energies by the
pure CS theories. Seiberg duality would then reduce to level-rank duality
in the corresponding WZW models giving a result that is inconsistent with the
$n$-dependent exchange $k\to nN-k$ provided by string theory.

At even larger coupling, the $U(k)$ theory is strongly coupled with a $\tr X^{n+1}$
deformation, but Seiberg duality provides a dual description in terms of a $U(nN-k)$
theory where the tree-level deformation is irrelevant and the IR theory is again 
controlled by the pure CS Lagrangian.

A more complete analysis of the dynamics of this theory with a verification of the
above scenario would be of interest.

\section{Conclusions}
\label{sec:conclusions}

In this paper we studied a class of $\NN=1$ supersymmetric Yang--Mills Chern--Simons 
theories in three dimensions  capturing the IR worldvolume dynamics of domain walls 
in the four-dimensional $SU(N)$ $\NN=1$ SYM theory and some of its generalizations. 
We argued that this class of CS theories can be reproduced as the low-energy dynamics 
on the worldvolume of D3-branes suspended between an NS5-brane and a $(1,N)$ 
fivebrane bound state in a type IIB string theory setup. By T-duality this setup resembles
the type IIA large-$N$ holographic dual of $\NN=1$ SYM theory presented in 
\cite{Vafa:2000wi}, but is not identical to it.

The brane configuration in type IIB string theory provides a simple intuitive 
understanding of some of the most characteristic properties of the $\NN=1$ 
SYM domain walls and the CSM theory that lives on them. These properties 
include: the identification of the $N-1$ BPS domain walls with supersymmetric
D3-brane configurations, a geometric counting of the Witten index and the
degeneracy of domain walls, and a strong/weak coupling Seiberg-type
duality of the CS worldvolume theory that relates the $k$-th wall with the
$(N-k)$-th anti-wall.

The brane construction is also a promising route for several
generalizations. Two of them were considered in section~\ref{sec:general}.
By adding D5-branes one adds $\NN=2$ flavor multiplets to the matter content
of the AV theory; by adding more NS5-branes one adds an $\NN=1$ scalar
multiplet in the adjoint with a tree-level superpotential. It would be interesting
to obtain a more complete understanding of the IR dynamics of the resulting
$\NN=1$ CSM theories. Moreover, it would be interesting to know if these
additional ingredients can also be embedded within a four-dimensional gauge
theory context.

An interesting open problem is to generalize the Acharya--Vafa theory to
the case of $\NN=1$ SYM with $SO$ (or $Sp$) gauge group. For this one
needs to consider a generalization of the brane setups in this paper that
includes the appropriate orientifolds. The resulting three-dimensional
worldvolume theory on the domain walls of these theories should admit again
a Seiberg duality which relates the $k$-th wall with the $(N-k-2)$-th wall
(or $(N-k+2)$-th wall).

Another generalization that presumably involves orientifolds is related to the
domain wall worldvolume theory of a four-dimensional $SU(N)$ gauge
theory with a Dirac fermion in the two-index anti-symmetric (or symmetric)
representation. Despite being a non-supersymmetric theory we expect
$N-2$ (or $N+2$) degenerate vacua, since the axial $U(1)$ is broken
by the anomaly to ${\Z}_{2(N-2)}$  (or ${\Z}_{2(N+2)}$) and then further to
${\Z}_2$. Moreover, it has been argued~\cite{Armoni:2003gp,Armoni:2003jk} that
in the large-$N$ limit this theory should admit `BPS' domain walls. Therefore,
we expect an Acharya--Vafa-like theory on its $k$-walls.

We conclude the paper by posing a question. We argued that the underlying reason
for Seiberg duality in the three-dimensional worldvolume theory is a four-dimensional
charge-conjugation symmetry. Is it possible that the original Seiberg duality for
$\NN=1$ SQCD in four dimensions~\cite{Seiberg:1994pq} is due to a discrete
symmetry in a higher dimensional theory?


\medskip
\section*{Acknowledgements}
\noindent

We would like to thank Ofer Aharony, Roberto Auzzi, Timothy Hollowood, Nissan Itzhaki,  
Nathan Seiberg and David Tong for discussions. The work of AA is supported
by the STFC advanced fellowship award.
The work of AG is supported in part by the BSF -- American-Israel Bi-National 
Science Foundation, by a center of excellence supported by the Israel Science 
Foundation (grant number 1468/06), DIP grant H.52, and the Einstein Center at 
the Hebrew University. AG thanks the EFI at the University of Chicago for a very 
warm hospitality.
The work of VN has been
supported by the European Union through an Individual Marie Curie Intra-European
Fellowship. Additional support was provided by the ANR grant,  ANR-05-BLAN-0079-02,
the RTN contracts MRTN-CT-2004-005104 and MRTN-CT-2004-503369, and the
CNRS PICS {\#} 3059, 3747 and 4172. VN thanks the organizers of the 
workshop ``New Perspectives in String Theory'' and the Galileo Galilei Institute, 
Florence for hospitality during the final stages of this work.

\end{document}